\documentclass[%
superscriptaddress,
showpacs,
nofootinbib,
amsmath,amssymb,
prc,
twocolumn,
floatfix ]%
{revtex4-2}

\usepackage{color}
\usepackage{CJK}

\usepackage{graphicx}
\usepackage{dcolumn}
\usepackage{bm}
\usepackage{hyperref}

\allowdisplaybreaks

\begin{document}

\begin{CJK*}{UTF8}{}
\title{Precise description of spontaneous fission half-lives of Rutherfordium isotopes 
	within the relativistic Hartree-Bogoliubov theory}
\CJKfamily{gbsn}
\author{Jie Zhao (赵杰)}%
\affiliation{Guangxi Key Laboratory of Nuclear Physics and Technology, Guangxi Normal University, Guilin, 541004, China}

\date{\today}

\begin{abstract}
I investigate the spontaneous-fission (SF) half-lives 
of $^{254,256}$Rf within the multidimensionally constrained 
relativistic Hartree-Bogoliubov (MDC-RHB) model combined 
with the local quasiparticle random-phase approximation (LQRPA) method. 
The collective inertia is evaluated using the finite-amplitude method (FAM), 
which enables a fully self-consistent treatment of dynamical residual effects 
beyond the conventional cranking approximation. 
The effective collective potential is constructed by subtracting 
both the vibrational zero-point energy and rotational energy correction 
from the deformation energy surface.
The SF half-lives are evaluated within the Wentzel-Kramers-Brillouin (WKB) approximation.
Compared with the 
	nonperturbative and 
perturbative cranking approximation, 
the FAM-LQRPA collective inertia is significantly enhanced, 
especially around the ground-state minimum, 
due to the dynamical residual interaction included in the QRPA. 
As a consequence, the predicted SF half-lives increase by several 
orders of magnitude and become consistent with experimental data. 
The calculated SF half-lives are 5.65 $\mu$s for $^{254}$Rf 
and 0.95 ms for $^{256}$Rf, both within approximately
	two orders
of magnitude of the corresponding experimental values. 
These results indicate that a self-consistent treatment of dynamical 
residual effects and beyond mean-field correlations is crucial 
for quantitative microscopic descriptions of SF within 
covariant density functional theory.
\end{abstract}

\maketitle

\end{CJK*}

\bigskip

\section{\label{sec:introduction}Introduction}
Accurate predictions of spontaneous-fission (SF) half-lives are 
essential for understanding the stability and synthesis of superheavy nuclei, 
as well as for modeling $r$-process nucleosynthesis in astrophysical environments.
Spontaneous fission is generally interpreted as a quantum tunneling phenomenon 
in a strongly correlated many-body fermionic system, 
whose fully microscopic dynamical description remains highly challenging. 
In most practical applications, SF half-lives are evaluated within 
the semiclassical Wentzel-Kramers-Brillouin (WKB) approximation for 
barrier penetration. 
The fission path embedded in a multidimensional collective space is usually 
determined either from the minimum-energy trajectory with respect to relevant 
collective coordinates, such as the quadrupole deformation $\beta_{20}$, 
or from the minimum-action principle. 
The key microscopic ingredients entering the WKB description are therefore 
the effective collective potential and the collective inertia along the fission path.

The collective potential can be calculated either within 
macroscopic-microscopic (MM) approaches 
or self-consistent mean-field frameworks based on 
effective interactions or energy density functionals (EDFs). 
In most modern studies, the collective inertia tensor is evaluated within 
the adiabatic time-dependent Hartree-Fock-Bogoliubov (ATDHFB) theory 
using either perturbative or nonperturbative cranking approximations. 
The perturbative cranking approximation neglects dynamical residual 
contributions and treats the derivatives of the densities with respect 
to collective coordinates perturbatively, 
whereas the nonperturbative approach explicitly evaluates these derivatives. 
Although perturbative cranking inertias have been widely employed 
in SF calculations~\cite{Brack1972_RMP44-320,
	Nilsson1969_NPA131-1,Girod1979_NPA330-40,Bes1961_NP28-42,
	Sobiczewski1969_NPA131-67}, 
a number of recent investigations~\cite{Baran2011_PRC84-054321,
	Sadhukhan2013_PRC88-064314,
	Zhao2015_PRC92-064315,Giuliani2014_PRC90-054311,
	Sadhukhan2014_PRC90-061304,Sadhukhan2016_PRC93-011304,
	Zhao2016_PRC93-044315,Giuliani2018_PLB787-134,Bernard2019_PRC99-064301} 
have demonstrated that nonperturbative 
treatments are crucial for a quantitative description of 
fission dynamics and SF half-lives. 
Systematic studies of SF lifetimes in superheavy 
nuclei have been carried out within both 
macroscopic-microscopic~\cite{Smolanczuk1995_PRC52-1871}
and self-consistent Hartree-Fock-Bogoliubov (HFB)
models~\cite{Baran2005_PRC72-044310,Staszczak2013_PRC87-024320}. 
In particular, recent calculations based on the nonrelativistic 
BSkG energy density functionals have achieved an overall agreement 
with experimental SF half-lives within less than four 
orders of magnitude~\cite{SanchezFernandez2026_PLB874-140287}.

An alternative framework for evaluating the collective inertia 
is based on the adiabatic self-consistent collective coordinate (ASCC) 
method~\cite{Matsuo2000_PTP103-959}. 
Within this approach, the collective subspace is generated 
from constrained HFB (CHFB) calculations, 
while the collective inertia is determined from local normal 
modes built on the constrained mean-field states. 
These local collective modes are obtained by solving the 
local quasiparticle random-phase approximation (LQRPA) equations. 
In contrast to the conventional cranking approximation, 
the dynamical residual interaction, including time-odd components, 
is self-consistently incorporated through the LQRPA
treatment~\cite{Washiyama2021_PRC103-014306}.

The CHFB+LQPRA framework has been successfully applied to the 
construction of quadrupole collective Hamiltonians and low-energy 
collective dynamics in finite 
nuclei~\cite{Hinohara2010_PRC82-064313,Sato2011_NPA849-53,
	Hinohara2011_PRC84-061302,Washiyama2024_PRC109-l051301,
	Sato2012_PRC86-024316,Hinohara2012_PRC85-024323,Yoshida2011_PRC83-061302}. 
More recently, by combining the momentum-coordinate (PQ) representation of 
the finite-amplitude method (FAM)~\cite{Nakatsukasa2007_PRC76-024318,
	Hinohara2015_PRC92-034321} with the LQPRA formalism, 
the CHFB+LQRPA approach has been extended to the calculation of 
collective inertia relevant for spontaneous 
fission~\cite{Washiyama2021_PRC103-014306,
	Washiyama2024_EPJWC306-01026}. 
The FAM implementation considerably reduces the computational cost of 
solving the QRPA equations and makes large-scale applications feasible. 
Previous studies based on nonrelativistic EDFs 
have shown that the resulting FAM-QRPA collective inertia is substantially 
enhanced compared with the perturbative cranking result, 
with pronounced peaks appearing around the ground-state minimum and fission isomer. 
These enhancements originate from the dynamical residual effects 
included in the QRPA correlations.

In recent years, covariant density functional theory (CDFT) 
has achieved considerable success in the description of potential-energy 
surfaces (PES's) and fission barriers of heavy and 
superheavy nuclei (Ref.~\cite{Lu2014_PRC89-014323} and references therein).
By allowing simultaneous breaking of axial and reflection symmetries, 
the multidimensionally constrained relativistic Hartree-Bogoliubov (MDC-RHB) 
models have been developed for the investigation of complex nuclear shape dynamics.
These approaches have been extensively applied to studies of fission barriers 
and PES's in actinide 
and superheavy nuclei~\cite{Lu2012_PRC85-011301R,Lu2014_PRC89-014323,
	Lu2012_EPJWoC38-05003,Lu2014_JPCS492-012014, Lu2014_PS89-054028,Lu2012_PhD,Zhao2015_PRC91-014321},
shapes of hypernuclei~\cite{Lu2011_PRC84-014328,Lu2014_PRC89-044307},
exotic nonaxial-octupole deformations~\cite{Zhao2012_PRC86-057304,
	Zhao2017_PRC95-014320,Zhao2024_PRC109-014303},
$\alpha$, $2\alpha$, and heavier cluster radioactivity~\cite{Mercier2020_PRC102-011301,
	Mercier2021_PRL127-12501,Zhao2023_PRC107-034311},
as well as spontaneous and induced fission dynamics~\cite{Zhao2015_PRC92-064315,
	Zhao2016_PRC93-044315,Zhao2019_PRC99-014618,Zhao2019_PRC99-054613,
	Zhao2020_PRC101-064605,Zhao2021_PRC104-044612,Zhao2022_PRC105-054604,
	Zhao2022_PRC106-054609,Tao2017_PRC96-024319}.
More recently, the MDC-RHB framework has been further extended to FAM
calculations of multipole response functions in nuclei 
with exotic deformations~\cite{Zhao2024_PRC110-L011301}.

In the present work, I implement the LQRPA approach for 
the calculation of collective inertia within the MDC-RHB model. 
Illustrative calculations are performed for $^{254,256}$Rf. 
By consistently including the vibrational zero-point energy (ZPE) 
and rotational energy correction (REC) in the effective collective potential, 
together with the dynamical residual effects in the collective inertia 
through the LQRPA treatment, the calculated SF half-lives are found to 
agree with experimental data within approximately two order of magnitude.
The paper is organized as follows. 
In Sec.~\ref{sec:model}, I briefly introduce the theoretical framework. 
The calculated PES's, collective inertias, and SF half-lives 
are presented and discussed in Sec.~\ref{sec:results}. 
Finally, a summary and concluding remarks are given in Sec.~\ref{sec:summary}.

\section{\label{sec:model}Theoretical framework}
The SF half-life is evaluated within the one-dimensional (1D) WKB approximation,
which describes quantum tunneling through the collective potential barrier.
The semiclassical tunneling probability associated with a parametrized trajectory 
$L(s)$ in the collective space is written as 
\begin{equation}
	P(L) =  \frac{1}{1+\exp[2S(L)]},
	\label{eq:prob}
\end{equation}
where the action integral is
\begin{equation}
	\label{eq:act_integration}
	S(L) = \int_{s_{\rm{in}}}^{s_{\rm{out}}} {1\over\hbar} 
	\sqrt{ 2\mathcal{M}_{\rm eff}(s) \left[ V_{\rm eff}(s)-E_0 \right] } ds.
\end{equation}
Here, $\mathcal{M}_{\rm eff}(s)$ and $V_{\rm eff}(s)$ denote the effective 
collective inertia and effective collective potential along the fission 
path $L(s)$ respectively.
The lower and upper integration limits, 
$s_{\rm{in}}$ and $s_{\rm{out}}$, are the classical inner 
and outer turning points determined by $V_{\rm eff}(s)=E_0$. 
The SF half-life is then obtained as $T_{1/2}=\ln2/(nP)$, where $n$ is the
assault frequency on the fission barrier~\cite{Baran1978_PLB76-8,
	Baran1981_NPA361-83,Sadhukhan2013_PRC88-064314,Sadhukhan2014_PRC90-061304}.
	
The effective inertia along the collective path is related to the 
multidimensional collective inertia tensor 
though~\cite{Brack1972_RMP44-320,
	Baran1978_PLB76-8,Baran1981_NPA361-83,Sadhukhan2013_PRC88-064314,
	Sadhukhan2014_PRC90-061304}
\begin{equation}
	\mathcal{M}_{\rm eff}(s) = \sum_{ij} \mathcal{M}_{ij} {dq_i \over ds} {dq_j \over ds},
\end{equation}
where $\mathcal{M}_{ij}$ denotes the components of the collective inertia tensor,
and $q_{i}(s)$ are the collective coordinates along the path.

In the present work, both the effective potential and the collective inertia 
are obtained from constrained RHB calculations.
The constrained calculations are preformed within the 
extend MDC-RHB model~\cite{Zhao2024_PRC109-014303,Zhao2024_PRC110-L011301}, 
in which the RHB equation
\begin{equation}
	\label{eq:rhb}
	\left(\begin{array}{cc} h_{D}-\lambda & \Delta \\ -\Delta^{*} & -h_{D}+\lambda \end{array}\right)
	\left(\begin{array}{c} U_{k} \\ V_{k} \end{array}\right)
	= E_{k}\left(\begin{array}{c} U_{k} \\ V_{k} \end{array}\right),
\end{equation}
is solved by expanding the Dirac spinors in a simplex-$y$ harmonic oscillator basis.
Here, $E_{k}$ is the quasiparticle energy, 
$\left( U_k, \ V_{k} \right)^\mathrm{T}$ is the quasiparticle wave function,
$\lambda$ denotes the chemical potential,
$h_{D}$ is the single-particle Dirac Hamiltonian,
and $\Delta$ is the pairing potential.
The particle-hole channel is described by the covariant 
density functional DD-PC1~\cite{Niksic2008_PRC77-034302},
while a separable pairing force of finite range is adopted in the 
particle-particle channel~\cite{Tian2009_PLB676-44, 
	Tian2009_PRC80-024313, Niksic2010_PRC81-054318}.
All calculations are carried out in a basis with 16 oscillator shells.
The deformation parameter $\beta_{\lambda\mu}$ is obtained from the corresponding
multipole moment as
\begin{equation}
	\beta_{\lambda\mu}^{\tau}=\frac{4\pi}{3N_{\tau}R^{\lambda}}Q_{\lambda\mu}^{\tau},
\end{equation}
where $R=1.2 \times A^{1/3}$ fm
and $N_{\tau}$ denotes the number of proton, neutron, or nucleons.

The effective collective potential $V_{\rm eff}$ is obtained by subtracting the 
vibrational ZPE and the REC from the total RHB deformation energy.
Within the Gaussian overlap approximation, 
the vibrational ZPE is calculated as~\cite{Staszczak2013_PRC87-024320,
	Baran2011_IJMPE20-557,Sadhukhan2013_PRC88-064314,Sadhukhan2014_PRC90-061304}, 
\begin{equation}
	\label{eq:zpe}
	E_{\rm ZPE} = {1\over4} {\rm Tr} \left[ {\it M}_{(2)}^{-1} {\it M}_{(1)} \right].
\end{equation}
with
\begin{equation}
	\label{eq:mmatrix}
	\left[ {\it M}_{(k)} \right]_{ij} = \sum_{\mu\nu} 
	{\left\langle 0 \left| \hat{Q}_i \right| \mu\nu \right\rangle
		\left\langle \mu\nu \left| \hat{Q}_j \right| 0 \right\rangle
		\over (E_\mu + E_\nu)^k},
\end{equation}
where $|\mu\nu\rangle$ are two-quasiparticle wave functions.
The REC is taken as
\begin{equation} 
	\label{eq:rec}
	E_{\rm REC} = \frac{<J^2>}{2\mathcal{I}_Y}\;,
\end{equation}
where the expectation values of the angular momentum operator 
are evaluated with respect to the self-consistent and 
deformation-constrained RHB states, 
and $\mathcal{I}_Y$ is the Yoccoz moment of inertia~\cite{Egido2004_EDFNSP}.
In the quasiparticle representation, it can be evaluated as~\cite{Girod1979_NPA330-40}
\begin{equation}
	\label{eq:yoc}
	\mathcal{I}_Y = 2\hbar^{2} \sum_{\mu\nu} 
	{\left|\left\langle 0 \left| \hat{J}_{Y} \right| \mu\nu \right\rangle\right|^2
		\over  E_\mu + E_\nu}\; . 
\end{equation}
The REC is expected to be reliable at large deformations~\cite{Warda2011_PRC84-044608}
and provides a practical approximation to full angular-momentum projection, 
which is computationally demanding for heavy nuclei.
The self-consistent RHB solutions, including 
quasiparticle energies and wave functions, provide the microscopic 
input for calculating both zero-point and rotational corrections.

To evaluate the collective inertia beyond the perturbative cranking
approximation, we employ the constrained RHB plus LQRPA method.
Following the idea of CHFB+LQRPA 
approach~\cite{Washiyama2021_PRC103-014306,Washiyama2024_EPJWC306-01026},
we assume that a set of $N$ collective coordinates $q_{i}$ ($i=1,2,\cdots,N$)
self-consistently determined with the ASCC method~\cite{Matsuo2000_PTP103-959}
can be mapped onto a set of $N$ collective variables $s_m$ ($m=1,2,\cdots,N$),
and that the relevant collective subspace can be approximated by constrained 
RHB states generated with suitable constraining operators $\hat{s}_{m}$. 
The constrained RHB state $\phi(s)$ is obtained from the variational condition 
\begin{equation}
	\delta \left\langle \phi(s) \left|\hat{H}_{M} \right|\phi(s) \right\rangle =0,
\label{eq:var}
\end{equation}
with the constrained Hamiltonian 
\begin{equation}
	\hat{H}_{M} = \hat{H} - \sum_{\tau=n,p} \lambda_{\tau} \hat{N}_{\tau} 
	- \sum_{m} \lambda_{m} \hat{s}_{m}.
\end{equation}
Here, $\lambda_{n,p}$ are the neutron and proton Fermi energies,respectively,
and $\lambda_{m}$ are Lagrange multipliers introduced to impose the constraints 
$\left\langle \phi(s) \left|\hat{s}_{m} \right|\phi(s) \right\rangle = s_{m}$.
At each constrained RHB state, the LQRPA equations
\begin{equation}
	\delta \left\langle \phi(s) \left| 
	[\hat{H}_{M}, \hat{Q}_{i}(s)] - \frac{1}{i} \hat{P}_{i}(s)
	\right|\phi(s) \right\rangle = 0,
\end{equation}
\begin{equation}
	\delta \left\langle \phi(s) \left| 
	[\hat{H}_{M}, \frac{1}{i} \hat{P}_{i}(s)] - \Omega_{i}^{2}(s) \hat{Q}_{i}(s)
	\right|\phi(s) \right\rangle = 0,
\end{equation}
are solved to determine the local coordinate and momentum generators, 
$\hat{Q}_{i}(s)$ and $\hat{P}_{i}(s)$.
The quantity $\Omega_{i}(s)$ denotes the eigenfrequency of the corresponding
local normal mode.

Once the relevant LQPRA collective modes are identified, 
the collective inertia tensor in the constrained collective variables
is obtained through the coordinate transformation 
\begin{equation}
	\mathcal{M}_{mn}(s) = \sum_{i} 
	\frac{\partial q_{i}}{s_{m}}  \mathcal{M}_{ij} \frac{\partial q_{j}}{s_{n}},
	\label{eq:mass}
\end{equation}
where $\mathcal{M}_{ij}$ is the collective inertia tensor 
associated with the collective coordinates $q_{i}$.
In order to fix the arbitrary scale of $q_{i}$, 
we set $\mathcal{M}$ to be unity, $\mathcal{M}_{ij} = \delta_{ij}$.
The derivatives in Eq.~\ref{eq:mass} are evaluated 
from the local momentum generators as 
\begin{equation}
	\frac{\partial{s_{m}}}{\partial q_{i}} = 
	\left\langle \phi(s) \left| 
	\left[\hat{s}_{m}, \frac{1}{i} \hat{P}_{i}(s) \right]
	\right|\phi(s) \right\rangle.
\end{equation}

Following Ref.~\cite{Washiyama2021_PRC103-014306}, 
the LQRPA equations are solved using the FAM.
In the quasiparticle basis, the FAM equations are
\begin{equation}
	\label{eq:fam}
	\begin{aligned}
		\left( E_{\mu} + E_{\nu} - \omega \right) X_{\mu\nu}(\omega) 
		+ \delta H_{\mu\nu}^{20}(\omega) &= - F_{\mu\nu}^{20}, \\
		\left( E_{\mu} + E_{\nu} + \omega \right) Y_{\mu\nu}(\omega) 
		+ \delta H_{\mu\nu}^{02}(\omega) &= - F_{\mu\nu}^{02}.
	\end{aligned}
\end{equation}
Here, $X$ and $Y$ are the FAM amplitudes at frequency $\omega$,
while $\delta H^{20}$ and $\delta H^{02}$ are the two-quasiparticle 
components of the induced Hamiltonian.
The quantities $F^{20}$ and $F^{02}$ are the corresponding two-quasiparticle 
components of the external field.
After convergence of the iterative FAM equations, 
the strength function is calculated as
\begin{equation}
	\label{eq:strength}
	S(\hat{F},\omega) = \sum_{\mu<\nu} \left[
	F_{\mu\nu}^{20*} X_{\mu\nu}(\omega) + F_{\mu\nu}^{02*} Y_{\mu\nu}(\omega)
	\right].
\end{equation}
The transition strength distribution is then given by
\begin{equation}
	\frac{dB(\hat{F},\omega)}{d\omega} = -\frac{1}{\pi} \rm{Im} S(\hat{F},\omega).
\label{eq:sdist}
\end{equation}
For the details of the formulas of the relativistic QFAM, 
we refer the readers to Ref.~\cite{Zhao2024_PRC110-L011301,Bjelcic2020_CPC253-107184}.

In the PQ representation of the QPRA~\cite{Hinohara2015_PRC92-034321},
the Hermitian coordinate and conjugate momentum operators,
$\hat{Q}_{i}$ and $\hat{P}_{i}$ are constructed from the QRPA phonon operators.
With the two-quasiparticle components $P^{i}_{\mu\nu}$ and $Q^{i}_{\mu\nu}$, 
the FAM amplitudes $X$ and $Y$ can be expressed as  
\begin{equation}
\begin{aligned}
	X_{\mu\nu}(\omega) =& \sum_{i} \frac{1}{\omega^{2}-\Omega_{i}^{2}}
	\left[ (-i\omega P^{i}_{\mu\nu} + \Omega^{2}_{i} Q^{i}_{\mu\nu}) q_{i}(\hat{F}) 
	\right. \\
	 &+ \left. (P^{i}_{\mu\nu} + i\omega Q^{i}_{\mu\nu}) p_{i}(\hat{F}) \right],
\end{aligned}
\end{equation}
\begin{equation}
	\begin{aligned}
		Y_{\mu\nu}(\omega) =& \sum_{i} \frac{1}{\omega^{2}-\Omega_{i}^{2}}
		\left[ (i\omega P^{i*}_{\mu\nu} - \Omega^{2}_{i} Q^{i*}_{\mu\nu}) q_{i}(\hat{F}) 
		\right. \\
		&+ \left. (-P^{i*}_{\mu\nu} - i\omega Q^{i*}_{\mu\nu}) p_{i}(\hat{F}) \right].
	\end{aligned}
\end{equation}
The quantities $p_{i}(\hat{F})$ and $q_{i}(\hat{F})$ are defined by
\begin{equation}
	p_{i}(\hat{F}) = \left\langle 0 \left| 
	\left[\hat{P}_{i}, \hat{F} \right] \right| 0 \right\rangle
	= \sum_{\mu\nu} (P^{i*}_{\mu\nu} F^{20}_{\mu\nu} - P^{i}_{\mu\nu} F^{02}_{\mu\nu}),
\end{equation}
\begin{equation}
	q_{i}(\hat{F}) = \left\langle 0 \left| 
	\left[\hat{Q}_{i}, \hat{F} \right] \right| 0 \right\rangle
	= \sum_{\mu\nu} (Q^{i*}_{\mu\nu} F^{20}_{\mu\nu} - Q^{i}_{\mu\nu} F^{02}_{\mu\nu}).
\end{equation}
Using these expressions, 
the FAM strength function Eq.~(\ref{eq:strength}) can be rewritten as 
\begin{equation}
	S(\hat{F},\omega) = \sum_{i} \frac{1}{\omega^{2} - \Omega_{i}^{2}}
	\left[ \left|p_{i}(\hat{F})\right|^{2} + 
	\omega_{i}^{2} \left|q_{i}(\hat{F})\right|^{2} +
	\omega r_{i}(\hat{F})
	\right],
\label{eq:strength1}
\end{equation}
where
\begin{equation}
	r_{i}(\hat{F}) = i \left[ 
	q_{i*}(\hat{F}) p_{i}(\hat{F}) - q_{i}(\hat{F}) p_{i*}(\hat{F}) \right]
\end{equation}
is a real quantity.

In the present study, we restrict the collective dynamics to a 
single collective coordinate associated with the isoscalar quadrupole degree of freedom.
The constraining operator is chosen as $\hat{s} = \hat{Q}_{20}$,
and the same operator is used as the external field $\hat{F} = \hat{Q}_{20}$.
With the coordinate transformation from the QRPA collective coordinate $q$ to
$s=	\left\langle \phi(s) \left| \hat{Q}_{20} \right|\phi(s) \right\rangle$, 
the one-dimensional collective inertia becomes 
\begin{equation}
	\mathcal{M}= 
	\frac{\partial q}{s} \frac{\partial q}{s}.
	\label{eq:mass1d}
\end{equation}
The derivative is evaluated as 
\begin{equation}
	\frac{\partial{s}}{\partial q} = 
	\left\langle \phi(s) \left| 
	\left[\hat{Q}_{20}, \frac{1}{i} \hat{P}_{i} \right]
	\right|\phi(s) \right\rangle
	= ip_{i}(\hat{Q}_{20}).
\end{equation}
Because the external-field operator $\hat{F}=\hat{Q}_{20}$ 
is real in the present case and its two-quasiparticle components are real,
one obtains $q_{i}(\hat{F})=0$, while $p_{i}(\hat{F})$ is purely imaginary. 
The strength function is therefore simplified to
\begin{equation}
	S(\hat{F},\omega) = \sum_{i} 
	\frac{1}{\omega^{2} - \Omega_{i}^{2}} \left|p_{i}(\hat{F})\right|^{2}.
\label{eq:strength2}
\end{equation}
Using Cauchy's integral formula, the residue associated 
with the pole $\omega=\Omega_{i}$ gives
\begin{equation}
	\frac{1}{2\pi i} \oint_{C_{i}} \omega S(\hat{F},\omega) d\omega 
	= \frac{1}{2} \left|p_{i}(\hat{F})\right|^{2}.
\end{equation}
Here, the contour $C_{i}$ in the complex energy plane is chosen 
to encircle only the pole at $\omega = \Omega_{i}$.
To determine the appropriate contour for each QRPA pole, 
we first calculate the FAM strength distribution with the 
isoscalar quadrupole external field $\hat{Q}_{20}$.
The frequency is taken as complex, $\omega = \omega_{R} + i\omega_{I}$.
For calculations along the real and imaginary axes, the imaginary
and real components of $\omega$, respectively, play the role of 
a smearing width, which is chosen as $1.0$ MeV.
The QRPA poles are identified from the sharp peaks in the FAM strength 
distribution below $\omega^{2}< 25$ MeV$^{2}$.
For each selected pole, $\left|p_{i}(\hat{F})\right|^{2}$ is extracted 
by contour integration along a circle centered at the estimated peak frequency.
The radius is taken to be $0.2$ MeV, and the contour is discretized 
with twelve points. 
The collective mode with the largest value of $\left|p_{i}(\hat{F})\right|^{2}$ 
is adopted as the most collective mode.

For comparison, the collective inertia is also evaluated using 
the nonperturbative and perturbative cranking approximation 
based on the adiabatic time-dependent HFB expression.
In the nonperturbative cranking approximation~\cite{Baran2011_PRC84-054321,
	Zhao2015_PRC92-064315,Zhao2016_PRC93-044315} 
\begin{equation}
	\label{eq:npmass}
	\mathcal{M}_{ij}^{C} = {\hbar^2 \over 2 \dot{q}_i \dot{q}_j}
	\sum_{\alpha\beta} {F^{i*}_{\alpha\beta}F^{j}_{\alpha\beta} + F^{i}_{\alpha\beta}F^{j*}_{\alpha\beta}
		\over E_{\alpha} + E_{\beta}},
\end{equation}
where
\begin{equation}
	\label{eq:fmatrix}
	{F^{i} \over \dot{q}_{i}}  
	=  U^\dagger {\partial\rho \over \partial q_{i}} V^* 
	+ U^\dagger {\partial\kappa \over \partial q_{i}} U^*
	- V^\dagger {\partial\rho^* \over \partial q_{i}} U^*
	- V^\dagger {\partial\kappa^* \over \partial q_{i}} V^*\;.
\end{equation}
$\rho$ and $\kappa$ are 
the corresponding particle and pairing density matrices, respectively.
In the perturbative approximation, the inertia tensor 
reads~\cite{Brack1972_RMP44-320,Nilsson1969_NPA131-1,
	Girod1979_NPA330-40,Bes1961_NP28-42,Sobiczewski1969_NPA131-67,
	Baran2011_PRC84-054321}
\begin{equation}
	\label{eq:pmass}
	\mathcal{M}^{Cp} = \hbar^2 {\it M}_{(1)}^{-1} {\it M}_{(3)} {\it M}_{(1)}^{-1}.
\end{equation}
This perturbative cranking inertia is used as a reference to 
assess the effect of the LQPRA treatment of collective dynamics.

\section{\label{sec:results}Results and discussions}


\begin{figure}[htb]
	\begin{center}
		\includegraphics[width=0.45\textwidth]{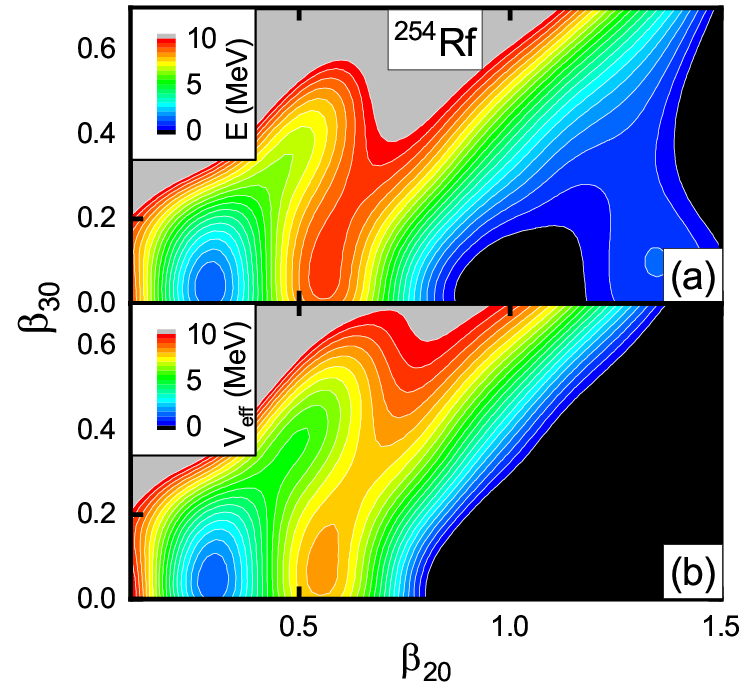}
	\end{center}
	\caption{(Color online) 
	Constrained self-consistent RHB energy surfaces (a) and effective collective
	potential obtained from the RHB energy by subtracting the zero-point vibrational 
	and rotational energy corrections (b) of
	$^{254}$Rf in the $(\beta_{20}, \beta_{30})$ plane.
	In each panel, energies are normalized with respect to the ground-state minimum, 
	and contours connect points of equal energy (in MeV). 
	The calculations employ the density functional DD-PC1~\cite{Niksic2008_PRC77-034302} 
	and a separable pairing interaction of finite range.
	Calculations are preformed assuming axial symmetry. 
	}
	\label{fig:Rf254pes}
\end{figure}

\begin{figure}[htb]
	\begin{center}
		\includegraphics[width=0.45\textwidth]{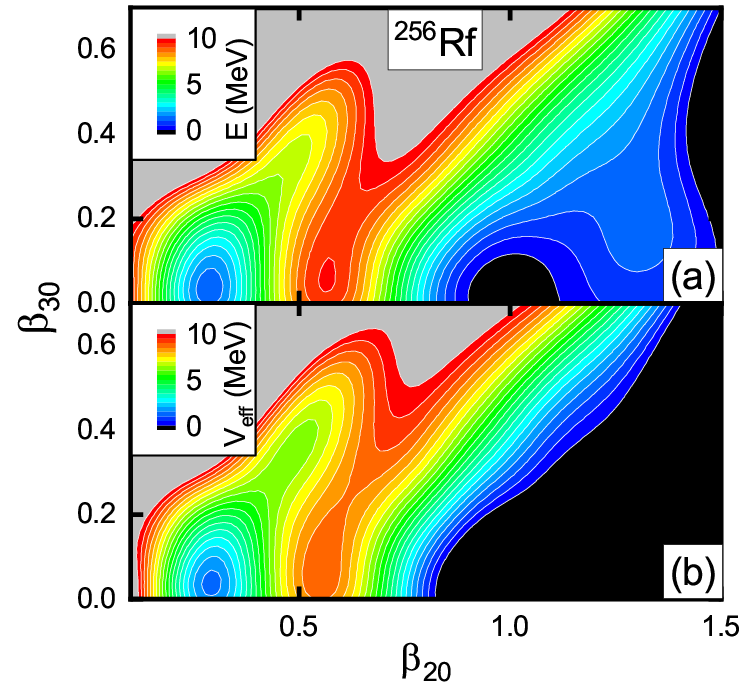}
	\end{center}
	\caption{(Color online) 
	Same as described in the caption to Fig.~\ref{fig:Rf254pes} 
	but for $^{256}$Rf.  
	}
	\label{fig:Rf256pes}
\end{figure}

\begin{figure}[htb]
	\begin{center}
		\includegraphics[width=0.45\textwidth]{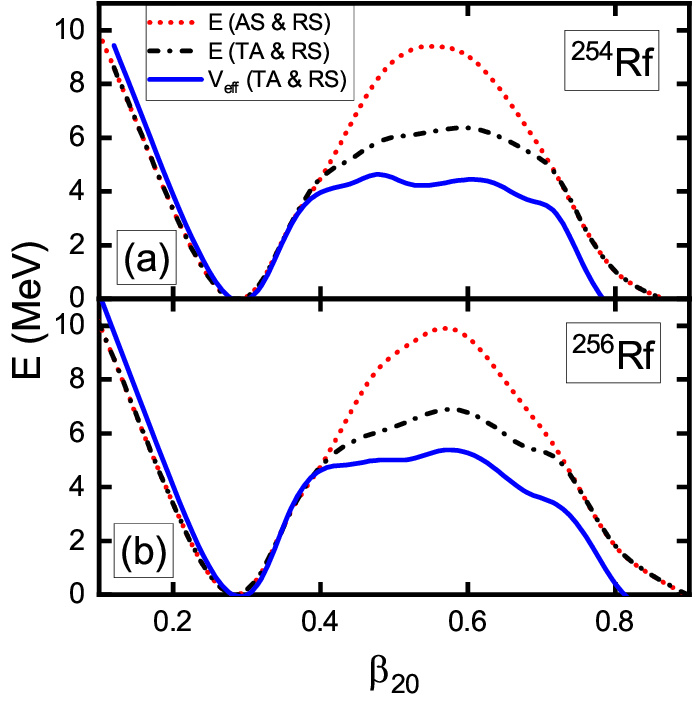}
	\end{center}
	\caption{(Color online) 
	Potential energy curves of $^{254}$Rf (a) and $^{256}$Rf (b)
	from RHB calculations. 
	The axially symmetric results are displayed by dotted curves,
	while those from the triaxial calculations are shown by dash-dotted curves.
	The effective collective potentials obtained from the RHB energy 
	by subtracting the zero-point vibrational and rotational energy corrections
	are denoted by solid curves.
	The energy is normalized with respect to the ground-state minimum of each nucleus.
	The calculations employ the density functional DD-PC1~\cite{Niksic2008_PRC77-034302} 
	and a separable pairing interaction of finite range.
	}
	\label{fig:pes1d}
\end{figure}

\begin{figure}[htb]
	\begin{center}
		\includegraphics[width=0.45\textwidth]{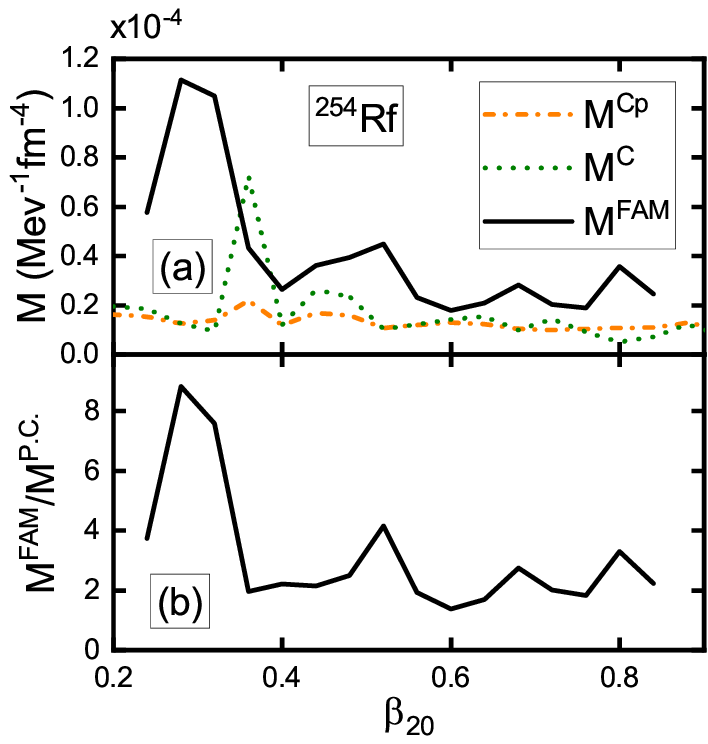}
	\end{center}
	\caption{(Color online)
	Collective inertia of $^{254}$Rf as a function of the quadrupole 
	deformation $\beta_{20}$ (a).
	The perturbative (nonperturbative) cranking inertia are shown 
	as a dash-dotted (dotted) curve, 
	while that from the FAM is shown as a solid curve. 
	The results are obtained from the triaxial calculations.
	The ratio of the FAM inertia to the perturbative cranking inertia
	is shown in panel (b).
	}
	\label{fig:Rf254mass}
\end{figure}

\begin{figure}[htb]
	\begin{center}
		\includegraphics[width=0.45\textwidth]{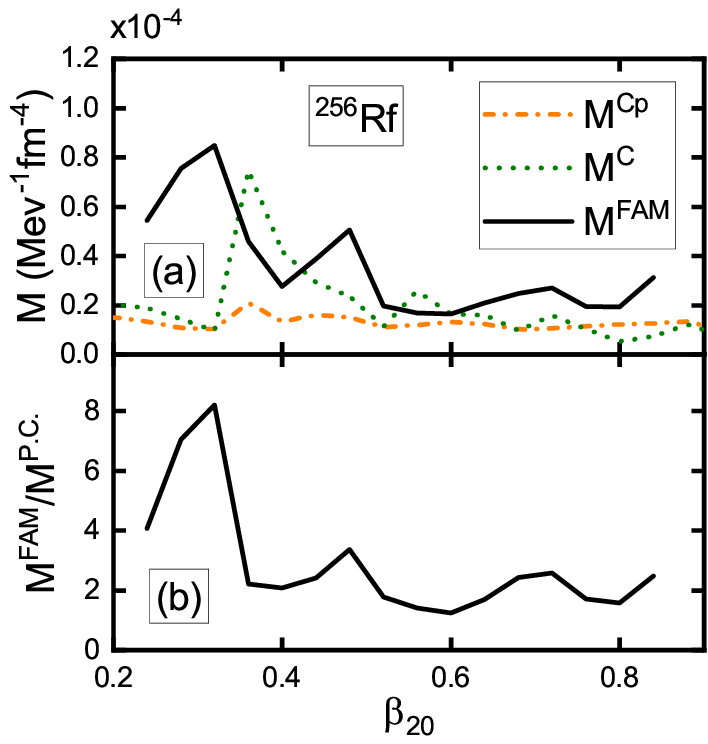}
	\end{center}
	\caption{(Color online) 
	Same as described in the caption to Fig.~\ref{fig:Rf254mass} 
	but for $^{256}$Rf.
	}
	\label{fig:Rf256mass}
\end{figure}

\begin{table}
	\caption{\label{tab:lives} %
		Half-lives for spontaneous fission of $^{254,256}$Rf,
		calculated with nonperturbative cranking ($T_{1/2}^{\rm{C}}$),
		perturbative cranking ($T_{1/2}^{\rm{Cp}}$),
		and FAM ($T_{1/2}^{\rm{FAM}}$) collective inertia.
	}
\begin{ruledtabular}
\begin{tabular}{lcccc}
Nucleus & $T_{1/2}^{\rm{Exp.}}$ & $T_{1/2}^{Cp}$ & $T_{1/2}^{C}$ & $T_{1/2}^{\rm{FAM}}$ \\ \hline
$^{254}$Rf ($\mu$s) & $22.8$    & $5.25\times 10^{-5}$  & $1.21\times10^{-3}$ & $5.65$    \\
$^{256}$Rf (ms)     & $6.76$    & $1.19\times 10^{-6}$  & $2.38\times10^{-3}$   & $0.95$    \\
\end{tabular}
\end{ruledtabular}
\end{table}

The heights of fission barriers are known to depend sensitively on 
pairing correlations~\cite{Karatzikos2010_PLB689-72}, 
and consequently the predicted fission dynamics can be 
significantly influenced by the choice of pairing strength. 
The finite-range separable pairing interaction adopted in the 
present work was originally adjusted to reproduce the pairing 
gap at the Fermi surface in symmetric nuclear matter obtained 
with the Gogny D1S force. 
Previous RHB studies have shown that additional fine tuning 
of the pairing strength is often necessary for a quantitative 
description of heavy and superheavy
nuclei~\cite{Wang2013_PRC87-054331,Afanasjev2013_PRC88-014320}.
In the present calculations, the pairing strengths for neutrons 
and protons are calibrated to reproduce the experimental odd-even
mass differences of $^{254,256}$Rf,
resulting in scaling factors $g_n/g_0=1.09$ and $g_p/g_0=1.08$, respectively.
Following our earlier studies~\cite{Zhao2016_PRC93-044315,Zhao2015_PRC92-064315,
	Mercier2020_PRC102-011301,Mercier2021_PRL127-12501,Zhao2023_PRC107-034311}, 
the collective ground state energy entering the action integral is taken as $E_0=1$ MeV.
Assuming a characteristic vibrational frequency $\hbar\omega_0=1$ MeV, 
the corresponding assault frequency is estimated to be $10^{20.38}$ s$^{-1}$.

Figure~\ref{fig:Rf254pes} (a) displays the self-consistent quadrupole-octupole 
deformation energy surface of $^{254}$Rf obtained from constrained RHB 
calculations with axial symmetry imposed. 
The mean-field ground state is predicted at a moderate quadrupole deformation of
$\beta_{20} \approx 0.3$, while a fission isomer appears at $\beta_{20}\approx1.0$.
The isomeric configuration is found to be slightly lower in energy than the
normal-deformed minimum by approximately 1 MeV. 
In the region of the inner fission barrier, 
the nucleus remains reflection symmetric, indicating that octupole deformation 
plays only a minor role in this part of the fission path. 
Beyond the fission isomer, however, reflection-asymmetric shapes gradually 
develop with increasing elongation. 
As a consequence, a shallow outer barrier emerges around $\beta_{20}\approx1.3$ 
and $\beta_{30}\approx0.3$. 
The height of the outer barrier is less than 1 MeV, 
and the potential-energy surface becomes rather soft in the vicinity of 
the outer saddle point.

The corresponding effective collective potential $V_{\rm eff}$, 
obtained after subtracting the vibrational ZPE and REC 
from the total deformation energy and normalized to the 
mean-field ground-state energy, is shown in Fig.~\ref{fig:Rf254pes} (b). 
Owing to the rapidly increasing rotational correction at large deformation, 
the effective potential becomes negative for $\beta_{20}\gtrsim0.9$. 
Since the calculation of SF half-lives only requires 
the classically forbidden region satisfying $V_{\rm eff}\ge E_0$, 
the tunneling dynamics is essentially governed by the inner barrier region.

To examine the inner barrier in more detail, the minimum-energy 1D
fission paths are shown in Fig.~\ref{fig:pes1d} (a). 
For the axially symmetric solution, shown by the dotted curve, 
the inner barrier height is predicted to be approximately 9.4 MeV. 
When triaxial deformation is included, 
the barrier is substantially lowered by about 3 MeV, 
as illustrated by the dash-dotted curve. 
The corresponding effective collective potential including triaxiality 
is shown by the solid curve. 
Compared with the total deformation energy, 
the effective potential exhibits a more complicated structure in 
the vicinity of the saddle point due to the deformation dependence 
of the vibrational ZPE and REC. 
In particular, both corrections vary rapidly around the triaxial 
saddle region and further reduce the effective barrier height 
relevant for the tunneling process.

In Fig.~\ref{fig:Rf256pes} (a), I display the self-consistent quadrupole-octupole
deformation energy surface of $^{256}$Rf obtained from constrained RHB 
calculations with axial symmetry imposed. 
Similar to $^{254}$Rf, the mean-field ground state is located at $\beta_{20}\approx0.3$,
while a fission isomer appears around $\beta_{20}\approx1.0$ 
and is predicted to be approximately 1 MeV lower in energy 
than the normal-deformed minimum. 
The inner barrier region remains nearly reflection symmetric, 
indicating a weak influence of octupole deformation.
Beyond the fission isomer, however, the deformation evolution 
differs from that obtained for $^{254}$Rf. 
The minimum-energy fission path continues to follow the reflection-symmetric 
axis with $\beta_{30}=0$, and no pronounced octupole instability develops 
along the outer fission path. 
A shallow outer barrier with a height of about 0.7 MeV is found 
around $\beta_{20}\approx1.3$. 
The corresponding effective collective potential $V_{\rm eff}$ 
is shown in Fig.~\ref{fig:Rf256pes}(b).
The minimum-energy 1D fission paths around the 
inner barrier are displayed in Fig.~\ref{fig:pes1d} (b). 
For the axially symmetric solution, the inner barrier height 
is predicted to be approximately 9.6 MeV. 
The inclusion of triaxial deformation lowers the barrier by about 2.8 MeV, 
resulting in an effective barrier height of approximately 6.8 MeV. 
The corresponding effective collective potential including triaxiality 
is shown by the solid curve. 
Similar to the case of $^{254}$Rf, the deformation dependence of the 
vibrational ZPE and REC produces a more intricate structure around 
the saddle region and further reduces the effective barrier relevant 
for spontaneous fission tunneling.

Another key ingredient in the evaluation of SF half-lives 
is the collective inertia tensor $\mathcal{M}_{ij}$. 
Although perturbative and nonperturbative cranking formulas 
based on the ATDHFB framework have been widely employed in previous studies, 
the dynamical residual interaction is not fully incorporated in these approaches. 
Similarly, the conventional GCM-GOA inertia~\cite{Giuliani2018_PLB787-134}, 
obtained from kernels between time-reversal-invariant constrained HFB states, 
does not explicitly include the complete dynamical 
time-odd, or Thouless-Valatin, response. 
Within the GCM framework, these contributions may be incorporated 
by extending the generator-coordinate manifold to include the 
collective momenta conjugate to the time-even coordinates. 
Alternatively, they can be treated through a self-consistent linear-response approach.
In particular, the recently proposed 
CHFB+LQRPA method~\cite{Washiyama2021_PRC103-014306,Washiyama2024_EPJWC306-01026} 
includes the dynamical residual effects explicitly through the local QRPA equations.
In the present work, I extend this framework to relativistic density 
functional theory and evaluate the collective inertia within 
the constrained RHB plus LQRPA approach.

As an illustrative example, Fig.~\ref{fig:Rf254mass} (a) displays the 
collective inertia of $^{254}$Rf in the vicinity of the inner barrier 
obtained from the constrained RHB+LQRPA calculations. 
The corresponding inertia is denoted by $\mathcal{M}^{\rm FAM}$. 
For comparison, the 
nonperturbative $\mathcal{M}^{C}$ 
and perturbative cranking inertia $\mathcal{M}^{Cp}$
is also shown. The ratio $\mathcal{M}^{\rm FAM}/\mathcal{M}^{Cp}$
is plotted in Fig.~\ref{fig:Rf254mass} (b). 
One observes that $\mathcal{M}^{\rm FAM}$ is systematically larger 
than the perturbative and nonperturbative cranking result over the entire deformation 
region considered here, with a pronounced enhancement around the 
ground-state minimum. This enhancement originates from the dynamical 
residual effects included in the LQRPA treatment. 
A qualitatively similar behavior has also been reported in previous 
nonrelativistic CHFB+LQRPA calculations based on Skyrme energy density
functionals~\cite{Washiyama2021_PRC103-014306,Washiyama2024_EPJWC306-01026}. 
The ratio $\mathcal{M}^{\rm FAM}/\mathcal{M}^{Cp}$ varies within 
the range of approximately 1.4-8.8, 
following the same deformation dependence as $\mathcal{M}^{\rm FAM}$. 
Such a large difference in the collective inertia has a substantial 
impact on the tunneling probability and consequently leads to several 
orders of magnitude variation in the predicted SF half-life.

The corresponding results for $^{256}$Rf are shown in Fig.~\ref{fig:Rf256mass}. 
The overall deformation dependence of the collective inertia is similar to 
that obtained for $^{254}$Rf, although the enhancement of $\mathcal{M}^{\rm FAM}$
relative to the perturbative cranking result is somewhat weaker, 
resulting in smaller peak values of the inertia ratio.

Using the calculated effective collective potential $V_{\rm eff}$ 
and collective inertia, we estimate the SF half-lives 
within the WKB approximation. 
For $^{254}$Rf, the SF half-life calculated using $\mathcal{M}^{\rm FAM}$ is $5.65~\mu\mathrm{s}$,
in reasonable agreement with the experimental value of $22.8~\mu\mathrm{s}$. 
In contrast, the perturbative cranking inertia yields a much shorter half-life 
of $5.25\times10^{-5}~\mu\mathrm{s}$, 
approximately five to six orders of magnitude below the experimental value. 
The nonperturbative cranking inertia gives $1.21\times10^{-3}~\mu\mathrm{s}$, 
which remains more than four orders of magnitude below the experimental value.
A similar trend is found for $^{256}$Rf. 
The half-life calculated using $\mathcal{M}^{\rm FAM}$ is 0.95 ms, 
within one order of magnitude of the experimental value of 6.76 ms. 
By comparison, the perturbative and nonperturbative cranking inertias 
yield $1.19\times10^{-6}$ ms and $2.38\times10^{-3}$ ms, respectively, 
both substantially shorter than the experimental value.
The calculated SF half-lives are summarized in Table~\ref{tab:lives}.

It is worth noting that REC are usually neglected in many previous 
EDF-based studies of SF pathways and
half-lives~\cite{Staszczak2013_PRC87-024320,
	Baran2011_IJMPE20-557,Sadhukhan2013_PRC88-064314,
	Sadhukhan2014_PRC90-061304,Zhao2015_PRC92-064315,
	Zhao2016_PRC93-044315,Washiyama2021_PRC103-014306,
	Washiyama2024_EPJWC306-01026}. 
However, the present analysis indicates that the REC plays a crucial 
role in obtaining realistic SF lifetimes. 
As an example, for $^{254}$Rf, neglecting the REC in the effective 
collective potential changes the predicted half-life from $5.65~\mu$s 
to $4.9\times10^{4}~\mu$s when $\mathcal{M}^{\rm FAM}$ is used, 
while the perturbative cranking result becomes $2.47\times10^{-2}~\mu$s. 
Both predictions deviate significantly from the experimental value, 
demonstrating the importance of including REC
consistently in the tunneling dynamics.

In addition, an accurate evaluation of SF half-lives requires an 
assessment of the uncertainties arising from basis truncation. 
Because evaluating the collective inertia within the LQRPA framework 
is computationally demanding, we estimate these uncertainties using 
perturbative cranking masses. 
When the number of oscillator shells is increased from 16 to 20, 
the fission barrier of $^{256}$Rf (including ZPE and REC)
is lowered by approximately 0.3 MeV, 
while the perturbative cranking masses decrease by approximately 7\% on average. 
Consequently, the calculated half-life decreases from $1.19\times10^{-6}$ ms 
to $0.15\times10^{-6}$ ms, corresponding to a reduction 
by a factor of approximately eight. 
The result obtained with 18 oscillator shells is $0.31\times10^{-6}$ ms. 
The smaller change from 18 to 20 shells, compared with that from 16 to 18 shells, 
indicates a trend toward convergence.
Because the FAM response is relatively insensitive to the size of 
the single-particle basis~\cite{Li2024_EPJWC292-10001}, 
we expect the basis-truncation uncertainty in $T_{1/2}^{\mathrm{FAM}}$ 
calculated with 16 oscillator shells to
remain within approximately one order of magnitude. 
After accounting for this uncertainty, 
the predicted $T_{1/2}^{\mathrm{FAM}}$ values agree with the 
experimental data within two orders of magnitude.

The present calculations are restricted to a 1D collective 
space in which the fission path is determined according to the minimum-energy 
principle. This simplification is mainly motivated by the extremely 
demanding computational cost associated with the evaluation of the 
collective inertia within the LQRPA framework. 
A fully multidimensional treatment that includes, 
for example, particle-number fluctuations as additional 
collective degrees of freedom~\cite{RodriguezGuzman2023_PRC107-044307,
	Zhao2016_PRC93-044315,Sadhukhan2014_PRC90-061304} 
and determines the fission trajectory according to the minimum-action principle 
will constitute an important extension of the present work. 
Moreover, the significant effect of the REC on the calculated SF half-lives 
highlights the importance of properly accounting for correlations associated with 
broken rotational symmetry. 
An explicit restoration of the relevant broken symmetries 
therefore represents another important direction for improving 
microscopic descriptions of SF.

\section{\label{sec:summary}Summary}
%
In this work, I have investigated the SF half-lives 
of $^{254,256}$Rf within the MDC-RHB model combined with the LQRPA method. 
The collective inertia was evaluated using the FAM within the LQRPA formalism, 
allowing the dynamical residual interaction, including time-odd components, 
to be incorporated self-consistently beyond the conventional cranking approximation.
Compared with the 
nonperturbative and 
perturbative cranking approximation, 
the collective inertia obtained from the FAM-LQRPA approach 
is systematically enhanced over the entire deformation region 
and exhibits pronounced peaks around the ground-state minimum. 
These enhancements originate from the dynamical residual effects 
included through the QRPA method. 
Consequently, the SF half-lives calculated with the FAM-LQRPA 
inertia are increased by several orders of magnitude relative 
to the 
	nonperturbative and 
perturbative cranking results and become much closer to 
the experimental data. 
The calculated half-lives of $^{254}$Rf and $^{256}$Rf are reproduced 
within approximately
	two orders
of magnitude.

The present study demonstrates that both the dynamical residual effects 
in the collective inertia and the beyond mean-field corrections in the 
effective collective potential are essential ingredients for a quantitative 
microscopic description of SF within CDFT. 
Although the present calculations are restricted to a one-dimensional 
collective space determined from the minimum-energy principle, 
the overall agreement with experimental data indicates that 
the MDC-RHB plus FAM-LQRPA framework provides a promising fully 
self-consistent approach for the description of SF dynamics in 
heavy and superheavy nuclei.
Future developments will include multidimensional calculations 
based on the minimum-action principle and systematic investigations 
of SF half-lives across the superheavy mass region.

\begin{acknowledgments}
This work has been supported by the National Natural Science 
Foundation of China under Grant No. 12005107.
\end{acknowledgments}


%

\end{document}